# Quantification of Biological Robustness at the Systemic Level

Aniket Magarkar, Anirban Banerji*, Shweta Kolhi

Bioinformatics Center, University of Pune, Pune-411007, Maharashtra, India.
Email address: **anirbanab@gmail.com**

**Abstract:**
Biological systems possess negative entropy. In them, one form of order produces another, more organized form of order. We propose a formal scheme to calculate robustness of an entire biological system by quantifying the negative entropy present in it. Our Methodology is based upon a computational implementation of two-person non-cooperative finite zero-sum game between positive (physico-chemical) and negative (biological) entropy, present in the system(TCA cycle, for this work). Biochemical analogue of Nash equilibrium, proposed here, could measure the robustness in TCA cycle in exact numeric terms, whereas the mixed strategy game between these entropies could quantitate the progression of stages of biological adaptation. Synchronization profile amongst macromolecular concentrations (even under environmental perturbations) is found to account for negative entropy and biological robustness. Emergence of synchronization profile was investigated with dynamically varying metabolite concentrations. Obtained results were verified with that from the deterministic simulation methods. Categorical plans to apply this algorithm in Cancer studies and anti-viral therapies are proposed alongside. From theoretical perspective, this work proposes a general, rigorous and alternative view of immunology.

**Introduction:**
Within every living cell, thousands of macromolecules interact in perfect harmony. Such harmony is displayed by the degree of synchronization within and between biochemical pathways. Synchronization ensures that new orders (with spectrum of scopes and depths) can continuously emerge from the existing paradigm of order with a non-trivial probability. We denote such a scheme, where one form of order may progressively generate more intricate form of order, by 'negative entropy'(NE). In stark contrast to this, the case of physico-chemical universe, within which the biological systems are embedded. The physico-chemical universe is characterized by the presence 'positive entropy'(PE). PE manifests itself with continuous random fluctuations in temperature, pressure, humidity, etc., and results in the production of disorder from existing order. The very fact that a crystal at 0K, shows perfect order but fails to create a morsel of richness in pattern that biological emergence can demonstrate - points clearly to an inherent difference in organization between objects of physico-chemical universe (computer, gas lighter, sky-scrappers etc.) and objects in biological universe (tree, fish, honey-bees etc.). This fundamental difference in organization, we argue, lies in the capability of production of order. Although entropy itself cannot assume a negative magnitude, since $p_i$ in the function ($-k_B \sum_i p_i \log p_i$) is positive and no larger than 1; NE refers to the negative (or in other words, less) production of entropy. To put it still differently, NE quantifies the probability of (spontaneous) production of more order from any existing template of order**.** By attributing the causality behind this spontaneous generation of order to the synchronization between metabolite concentrations present in a biochemical pathway (or a system of pathways), we propose here a scheme to calculate the robustness of an entire biological system. Present study fulfills all the three criteria, laid down by Kitano*(1)*, for being considered as the "Grand Unification Theory" of biological robustness.

We argue that to quantify biological robustness in numeric way at a systemic level, one needs to quantify the negative entropic content(NEC) of that system. In the present work NEC was measured by objectively describing the synchronization profile(SP), emerging from the coupling schemes in a biochemical pathway. Measurement of NEC was achieved by quantifying the amount of PE that is necessary for the system to discard its biological nature and resemble a physico-chemical system. This, in turn, was achieved by resorting to an implementation of two-person non-cooperative finite zero-sum game with pure strategy and mixed strategy, between positive (physico-chemical) and negative (biological) entropy. Equilibrium point of such an antagonistic game between PE and NE unambiguously identified the edge of life(EOL) for the biological system under consideration. EOL describes the profile of macromolecular concentrations of a biological system, when it resembles physico-chemical system. On the other hand, from the reference frame of physico-chemical system, EOL can be described as the profile of macromolecular concentrations with the rudimentary template that can possibly account for limited SP. The efficiency, with which any biological system adjusts itself to the PE of physico-chemical universe around it, is described by biological adaptation. Phenomenon of adaptation describes the capability of biological systems to re-organize their filtering mechanism, so as to fend off the physico-chemical (positive) entropy from entering and mixing with the biological negative) entropy. This filtering scheme translates to game theoretic parlance as 'optimal strategies'*(2)*.

Idea of NE isn't new; sheer relevance of this concept in the paradigm of biology was pointed out by Boltzmann*(3)*, Schrödinger*(4)*, Lovelock*(5)* amongst many others, in somewhat tangential and qualitative ways (**Supp Mat-2** contains somewhat detailed historical perspective and associated discussions). The present work assumes importance because of its capability to calculate the NE, for the first time. On the other hand, while there are many works on biological robustness, most of them have banked upon systematic perturbation analysis based studies of mere identification of the relevant concentration ranges of individual metabolite concentrations, changes in which can disrupt a pathway. In contrast to these, the composite, formal and general scheme proposed here quantified Biological Robustness at systemic level and also could bring to fore the causality behind its emergence.

Work was conducted on TCA cycle(TCAC). Although TCAC merely by itself cannot produce aforementioned SP, it is arguably the most conserved biochemical unit along the course of evolution. Such remarkable conservation hinted at the possibility of its possessing an enormous NEC. NEC of TCAC was measured across all the states by systematic and iterative measurement of difference in its magnitude between any state of the system and that in the thermodynamic equilibrium. This NEC of TCAC was found to be of the order of $10^{37}$ units (details are discussed under 'Results', section 2.2). Results obtained were verified with that from prevalent domain of deterministic differential equation based analysis of biological systems. This proved that the present work, in its highly reduced scope, conforms to the results obtained from existing realm of methods.

**Results and Discussion:**
Framework of internal dynamics of TCAC, viz. concentration of metabolites was represented by matrix-**A.** A typical physico-chemical system was modeled with matrix-**B** (matrix of random numbers, scaled within the range of **A**). Primary determination of depth of local and global NE minima on TCAC phase space, was achieved with pure strategy game between **A** and **B**. While this provided a rudimentary description about biological robustness **(**details in **Supp Mat-2)**; finer aspects of it was obtained from results of mixed strategy game.

Pure strategy game was played with uniform perturbation scheme, which introduced constant amount of PE and NE to biological and physico-chemical systems respectively. In contrast, in the mixed strategy game, players changed their perturbation schemes dynamically in response to opponent's strategy, to maximize respective gains. Pure strategy game between players (**A**, the biological system with NE; and **B**, the physico-chemical environment with PE) provided an objective estimate of robustness that a biological system can demonstrate, in order to keep itself functioning, under steady pattern of external disorderliness. Pure strategy game could further quantify how rapidly, **B** (PE of the physico-chemical universe), under NE perturbations, could change from its random nature and to a minimally disordered state. Concentration profile analysis showed that this minimally disordered state of the physico-chemical system resembles the profile of a biological system, to some extent (details can be Figure-1 of **Suppl. Mat-2**).

Game with mixed-strategy, on the other hand, provided the description of dynamic nature of biological system. In mixed strategy game, to ensure that emergence of SP is maintained; **A** ascertains the strategy of **B** at every instance to change the strategy of its own. This scheme could model TCAC's response to maintain and propagate SP under dynamic perturbations scheme, and hence, could quantify the stages of biological robustness with respect to external PE.

**Section – 1: Results obtained from game with pure strategies:**
**Section 1.1: Quantifying biological robustness by measuring the depth of negative entropic minima:**
At t=1 to t=2 the profile of (|A|) **(Figure 1A)** depicts TCAC's approach towards asymptotic stability. In other words, this stage described TCAC's approach to a state that is characterized by a favorable combination of concentration ranges of interacting metabolites (described by (|A|) and (|$A_i$|)). Presence of such a template of concentration ranges accounts for the emergence of nascent SP; which in turn, accounts for the emergence of local minima of nascent NE. Emergent SP attempted to nullify PE perturbations, as the TCAC approaches asymptotic stability. However, the nascent SP could merely account for a shallow local minima of NE, and therefore, it took only 32 PE perturbations to disrupt the nascent SP and take TCAC to the EOL **(Figure 1A Supp mat 2)**. At (t=3) the asymptotic stability of the system approached a deterministic limit and the depth of the NE local minima deepened. Consequently, it became difficult for the PE perturbations (applied to $A_i$) to drag TCAC out of its present depth of NE minima. Hence 74 PE perturbations were required to disrupt the SP. Presence of 42 (= $74_{t=3} - 32_{t=1,t=2}$) levels of additional PE, implied that the robustness of TCAC against PE is increasing. At the steady state (Supp Mat 2 **Figure 1D**, **Figure 1E**) robustness of TCAC was measured to be maximum, whereby it could account for large magnitude of PE

perturbations (115, more than 80 fold of PE perturbations needed to disrupt the nascent SP). One can generalize this finding to assert that the robustness of a biochemical pathway or network of biochemical pathways, reaches its highest magnitude at its steady state.

By alternatively describing the minimas of NE as attractors in the phase-space of TCAC, one can assert that the attractor, which during t=1 and t=2 could merely account for asymptotic Lyapunov convergence, matures into a stable fixed-point attractor, during steady state.

**Section 1.2: Biological mechanism to ensure NE minima:**

**Figure-1** shows comparative scaling of concentrations of systemic components. 'Number of metabolites' versus 'concentration ranges' for **B** showed a 'normal'(like) distribution of metabolites at time t=0, where every concentration bin was populated with similar probabilities (**Figure 1B**). However, with the introduction of NE, only a certain concentration range showed an increasing tendency of being populated. Finally, at Nash equilibrium, the concentration range acquired a 'power law'-(like) nature. For **A**, on the other hand, 10 out of 12 metabolites had their concentrations in a particular range of concentration (($5.27 \times 10^{10}$ to $1.05 \times 10^{10}$)μM) and only 2 could be observed to possess different concentration ranges than these 10 (**Figure 1A**). Even under PE perturbations, this heavily asymmetric pattern (10 metabolites in one concentration range, whereas 2 are scattered over 11 other possible ranges of concentration) persisted. At EOL of TCAC, the distribution of remained unchanged (with a mere range transformation). This demonstrated an unflinching tendency of TCAC to maintain most of its components in one concentration range (even under intense PE perturbations). Comparable ranges of concentrations probably helps in maintaining a viable platform to ensure continuation of coupled interactions, so that SP can emerge.

**Section 1.3: The Systematic Perturbation Analysis(SPA):**
In this analysis, individual metabolite concentration (barring the boundary metabolites) was varied, keeping rest of the metabolite concentrations constant. Actual biological analogue for such change could be mapped to events like that of gene knockouts, or the addition of a protein inhibitors, like a drug. With the SPA, role of every metabolite of TCAC in ensuring aforementioned SP was investigated. Since every metabolite's contribution to the change of entropy in the system could be quantified, the emergence and loss of SP for every metabolite could be obtained. (Details in **Supp Mat 2, Table-S1**).

**Section-2: Results obtained from game with mixed strategies:**
Here, NE and PE competed with dynamic strategies; that is, both attempt to maximize their gains by changing their strategies according to the last adopted strategy of the opponent. This is more reliable depiction of biological response; because, instead of responding to an unvarying PE perturbation scheme (pure strategy game), NE adjusts itself against continuously changing strategies of PE perturbations. This game, therefore, involved stages where players learnt from each other's strategy. **Figure 2A** and **Figure 2B**, depict the progression of stages of response of both biological and physico-chemical systems, under dynamically changing perturbations.

**Section-2.1: TCA cycle response to dynamic PE perturbations:**
**Figure 2A** shows how TCAC responded to dynamically changing PE, to ensure the SP. The learning phase of 76 interactions indicates **A**'s learning all the (dynamically changing) perturbation strategies of **B** and vice-versa. After this stage, both played with a common strategy. Consistent low magnitude of A's profile during first 76 interactions and beyond, described nascent adaptation of strategies of a biological system to protect its order against external PE. It hinted at the global depth of NE minima in TCAC phase-space.

**Section-2.2) SP, NE, Emergence of life - under dynamically changing PE perturbations:**
In **Figure 2B**, below the 'EOL', TCAC retains its functional (biological) state. Upon entering EOL, it could merely retain the concentrations of metabolites in such ranges where a primitive template SP of the pathway might exist. Such limited SP could only account for an instance of ordered interaction, lasting for a tiny interval of time. This transient order would fail to propagate more order and is in contrast to possibility of generation of order from order (NE), observed in 'Core of biological realm'. Close observation reveals that at the face of mounting PE, TCAC opts for a final effort to maintain the SP, before becoming a physico-chemical system with no SP. It is exactly through the same point in the state-space of the process, that the physico-chemical system acquires an ordered nature in the concentration profile of its constituents. However, although it acquires the SP, a ~40 fold logarithmic magnitude of determinant prevents it from acquiring the characteristics to support NE and Biological characteristics.

Huge difference (order of magnitude ~ $10^{37}$) between |**A**|(in 'core of biological realm') and |**B**|(with its highest ordered state). This difference measured the NEC necessary to propagate order from existing order. (Particular magnitudes of NEC and PE are dependent upon the applied perturbation scheme; a different perturbation strategy might yield different magnitude NEC and PE, keeping the general nature of the result invariant.)

**Section-2.3: Confirmation of game-theoretic results by COPASI and ODEs**

Game theoretic framework to study biochemical synchronizations could describe the emergence of SP and evolution of NE, but it failed short of understanding the internal dynamics of the system. To verify the results obtained from the top-down game-theoretic framework with a bottom-up ODE-centric framework, analysis was performed with COPASI(6). COPASI results confirmed the steady state of TCAC in, what has been described in this work as, the 'core of Biological realm'. More importantly, the state of physico-chemical system in "EOL" was ascertained to be "asymptotically stable". The ODE-based time course simulation of the biological system at different stages, as shown in Figure-3B, supported the present work from an all-different perspective.

**Theoretical implications:**

COPASI results (**Figure 3A**) vindicated the findings of **r**esults from concentration profile of NE minima (Fig- 1.A); both showed that in the NE dominated 'core of biological realm' region of a functional biological system, 10 out of 12 metabolites of TCAC reside in the same concentration range. Symmetrically, **Figure 3B** confirmed the findings of **Figure 2B**, in demonstrating a wide range of possible concentrations for physico-chemical entities, even though they are scaled within biologically observed range and even though they are subjected to the same system of interconnected differential equations that the TCAC metabolites are subjected to. **Figure 3** demonstrates that SP in EOL zone can at best account for a probabilistic occurrence of an ordered state; but merely that much falls short of ensuring the robust mechanism that the biological systems require to consistently minimize the effects of various PE perturbations. We assert that NE ensures this robustness; hence 'core of biological realm' stands out as the only zone of TCAC phase-space where order→order transitions can consistently be irreversible. Ironically, it seems that maintenance and generation of aforementioned order is achieved by allowing various components of the systems to fluctuate within certain permissible limits. Presence of almost all the metabolites in the same concentration range(**Figure 1A**), helps in achieving this. It is through the mappings between bounded magnitudes of parameter ranges, that, the consistent order→order scheme is maintained. For example, for mitogen-activated protein kinase cascades, total concentrations of MKKK, MKK and MAPK were found to be in the range 10–1000 nm and the estimates for the kcat values of the protein kinases and phosphatases were found to range between 0.01 to 1 *(7)*. The present study shows, how, these fluctuations do not go against ensuring the biological order and biological robustness, but helps in the construction of NE minima generated from SP in the face of environmental perturbations; and therefore, ensures biological robustness.

Kitano*(1)* had laid down three basic characteristics of the "Grand Unification Theory" of biological robustness. These are, a "solid quantitative index", an inherent ability to describe transitions between attractors when the trajectory of the system roams all across its phase-space to counter the effects of ever-changing perturbations; and third, avoidance of simplistic thermodynamic constructs. The present model satisfies all three of them in unambiguous ways. Amount of entropic shift (Fig-3) between NEC and PE, fulfills the criteria of a "solid quantitative index"; we assert that biological robustness is directly proportional to it. Locus of NEC trajectory, when being perturbed with PE across all-possible attractors in its phase-space (Fig-3) fulfills the second criterion. Since thermodynamics has been touched in the present model more from a conceptual angle than from the perspective of rigorous modeling, the third is satisfied too.

**Practical implementation:**

Cancer is a typical example of 'robustness hijacking'*(1)*. Here, the SP of biochemical pathways undergoes such local or a global change(s) that a new form of more robust SP emerges, which shields them from known therapeutic strategies. For example in VEGF pathway, HIF-1 up-regulates VEGF and series of genes that in turn triggers angiogenesis and cell motility. HIF-1 up-regulation takes place because of hypoxia in tumors and dysfunctional blood vessels caused by tumor growth. This feedback regulation enables tumor growth and metastasis. However, HIF-1 up-regulation is important for normal physiology under oxygen-deprived conditions *(1)*. This indicates that normal physiological mechanisms are effectively hijacked. Present approach ensures as exact identification of the point of emergence of such robust SP, both globally and locally. SPA measures unambiguously the contribution of each metabolite to the changing SP. It can be applied for the modern Anti-viral therapeutic strategies too.

***Acknowledgment:*** *Authors would like to acknowledge COE-DBT(Govt. of India) scheme, Dr. Urmila Kulkarni-Kale, Professor Indira Ghosh and Professor Ashok Kolaskar, for supporting them during the tenure of this work.*

**Materials and Methods :**

**Section 1 :**
**Basic (mathematical) framework :**
TCAC concentrations for the first 100 minutes [8] was described by 88 matrices (12X12) and served as the template of description of (pure and mixed) strategies for NE calculation in it. 88 matrices of random numbers scaled within TCAC metabolite concentrations, described the (pure and mixed) strategies of PE representing the (external) physico-chemical environment.

Scheme for such a game was constructed with simple equation $G = <A, B, H>$, where **A** and **B** denote the sets of possible actions of them, respectively; while **H** denotes the utility function for **A** defined on **AxB**. While **A** forms the set of 'pure strategies' of NE, **B** represents the set of 'pure strategies' of PE, as H represented the pay-off function for **A**. More clearly, **A** and **B** were described as :
$$\mathbf{A} = \begin{bmatrix} a_{11} & a_{12} & \cdots & a_{1n} \\ a_{21} & a_{22} & \cdots & a_{2n} \\ \cdots & \cdots & \cdots & \cdots \\ a_{m1} & a_{m2} & \cdots & a_{mn} \end{bmatrix} \text{ and } \mathbf{B} = \begin{bmatrix} b_{11} & b_{12} & \cdots & b_{1n} \\ b_{21} & b_{22} & \cdots & b_{2n} \\ \cdots & \cdots & \cdots & \cdots \\ b_{m1} & b_{m2} & \cdots & b_{mn} \end{bmatrix}$$

Number *m* and *n* in matrices **A** and **B**, describe pure strategies of (*negative entropy*) and (*positive entropy*), respectively. From biological perspective, any particular element, say $a_{ij}$ for the matrix **A**, denotes the concentration of some macromolecule in matrix **A**; whereas for the game-theorists it implies the payoff of (*negative entropy*) in any situation, when being subjected to strategies *i,j*.

Published biological data [8] for concentrations of 12 metabolites for the first 100 minutes of evolution of TCAC was considered as the pure strategy of NE described by **A**. PE was represented in the same framework (identical to **A**) by a 12x100 matrix, **B**; where all the entries were random numbers. It was assumed that probability of a synchronization profile emerging out of a set of random numbers is zero, for all practical purposes. However, to ensure unbiased comparisons and a rapid convergence of the algorithm (by negating the scope of absurd comparability), generated random numbers were scaled within the same range of concentration as that of the entries of **A**.

The Nash equilibrium (captured by the pair of strategies $(X^*, Y^*)$) for this bi-matrix game exists if :
$(X^* A Y^{*T}) \geq A_{i \rightarrow} Y^{*T}$ (where i = 1, 2, …, m)     (Eq$^n$-1)
and
$(X^* B Y^{*T}) \geq X^* B_{\rightarrow j}$ (where j = 1, 2, …, n)     (Eq$^n$-2)

Here $(X^* \geq 0)$ and $(Y^* \geq 0)$ are optimal strategies which ensure payoffs to both adversaries; $B_{\rightarrow j}$ (denoting row of **B**) and $A_{i \rightarrow}$ (denoting row of **A**). The composite framework can therefore be expressed as :
$(X^* E_m^T) = (E_n Y^{*T}) = 1$ ,     $E_\psi = (1,1,…,1)$ ™ $R^\psi$     (Eq$^n$-3)

**Section 2 :**
**Computational Implementation:**
Computational implementations (and validations) were performed with MATLAB and COPASI. Only the significant points are mentioned below (details of these can be found in **Supp Mat-1**).

**Section 2.1): Implementation of pure strategy game (between biological system and physico-chemical environment, under invariant schemes of PE and NE perturbations, respectively):**

**2.1.1: Perturbation of A (biological system), pure strategy:**
12 metabolites were sorted with respect to their initial concentration (t=1). First 3 metabolites of these 12 were identified as the boundary metabolites; they were kept constant throughout the course of simulation unless the system was not at steady state. Applying random fluctuations to ensure PE perturbations to biological system was difficult for the construction of an invariant and formal framework. Hence, metabolites were perturbed with proportional perturbation scheme. Here, concentrations of metabolites with low magnitudes (total 4 of them) were increased by 50% of their individual magnitudes at that instance; while, concentrations of metabolites with high magnitudes (total 5 of them) were decreased by 50% of their individual magnitudes at that instance. This scheme of iterative perturbations was continuously applied to each metabolite of **A** during the entire simulation course of pure strategy game. Proportional perturbation helped in SPA, because patterns of any particular metabolite's ability to affect the entire system of synchronized concentrations could be studied with it at a formal and systemic scale.

**2.1.1: Perturbation of B (physico-chemical system), pure strategy:**
While the biological system (TCAC, described by **A**) was being perturbed with PE, the physico-chemical environment (modeled by system of random numbers, described by **B**) was being perturbed with NE. The later was achieved by an equivalent scheme where PE was continuously depleted from **B**. For the case $(([b_{ij}] - [a_{ij}]) > 0)$, $[b_{ij}] = [b_{ij}] - (([b_{ij}] - [a_{ij}])/2)$ was (iteratively) implemented; whereas for the case $(([b_{ij}] - [a_{ij}]) < 0)$, $[b_{ij}] = [b_{ij}] + (([b_{ij}] - [a_{ij}])/2)$ was (iteratively) implemented. It is easy to notice that merely subtracting the (fictitious) concentrations from **B** will not account for the reduction in its entropy profile, but only bring the (a) synchronization-profile to a lower magnitude.

**Section 2.2: Implementation of mixed strategy game (between biological system and physico-chemical environment, under dynamically varying schemes of PE and NE perturbations, respectively):**

To implement the dynamically varying strategies of both players, the biological system (**A**) was made to acquire 10% of physico-chemical system (**B**)'s characteristics and vice versa, for all the iterations during the entire simulation. Plot of (|**A**|) and (|**B**|) showed evolution of magnitudes of them.

In order to simulate $eq^n$-1 and $eq^n$-2, magnitudes of **A**, **B**, **X**, **Y**, **X**$^T$ and **Y**$^T$ were calculated with precise numerical magnitudes. The entire range of concentrations was divided in 12 bins of concentrations. Each of these bins represented a (possible) range of concentration of metabolites of TCAC. This meant that each one of **A**, **B**, **X**, **Y**, **X**$^T$ and **Y**$^T$ was represented as 1-dimensional matrices. Range of possible concentrations was represented by A, the number of metabolites with a given range of concentration was represented by X; and **X**$^T$, denoted the transpose of **X**. (**B**, **Y** and **Y**$^T$ was represented accordingly).

**Section 2.3: Validation by COPASI and ODEs**

In order to validate the results obtained from (top-down) game-theoretic scheme, (bottom-up) ODE-centric framework was resorted to. Systemic characteristics of different stages of biological system, as shown in Figure-3, were analyzed by COPASI-based steady state analysis (with default software parameters). Alongside this, in order to study the dynamics of the system at different stages viz. EOL, gain and loss of SP, core of biological realm; analysis was performed with ODEs in 'Matlab'.

Details of the mathematical framework and algorithmic implementation schemes are elaborated in **Supp Mat-1**.

Figure-1.A:
Title: Metabolite distribution of **A** over time

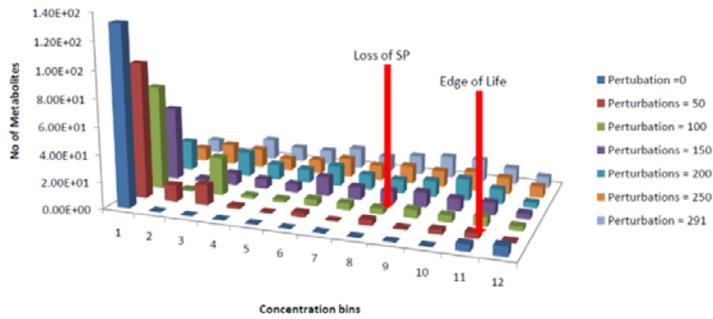

**Fig-1.A legends:** Skewed nature (distribution of metabolites in different concentration bins remains almost the same till EOL) implies that majority of metabolites in TCA cycle tend to reside in particular concentration ranges; it helps ensuring a feasible platform for emergence of SP (and NE).

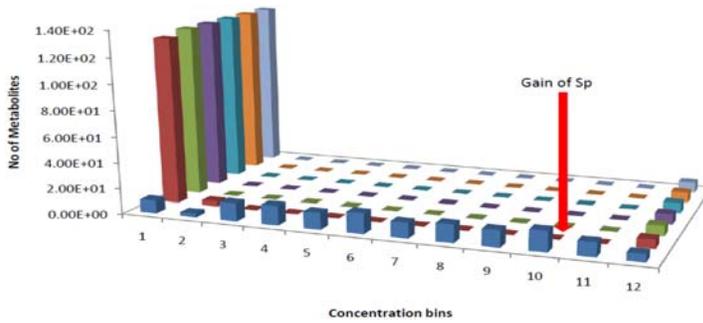

**Fig-1.B legends :** Distribution of (purported) metabolites in Physico-Chemical system before and after introduction of NE. In the mixed strategy game, **B** starts to acquire **A**'s nature, certain concentration bins show tendencies of being more populated, instead of every bin being populated with similar probabilities.

**Fig-2A**: Title: How NE resists (dynamic) PE and ensures robustness

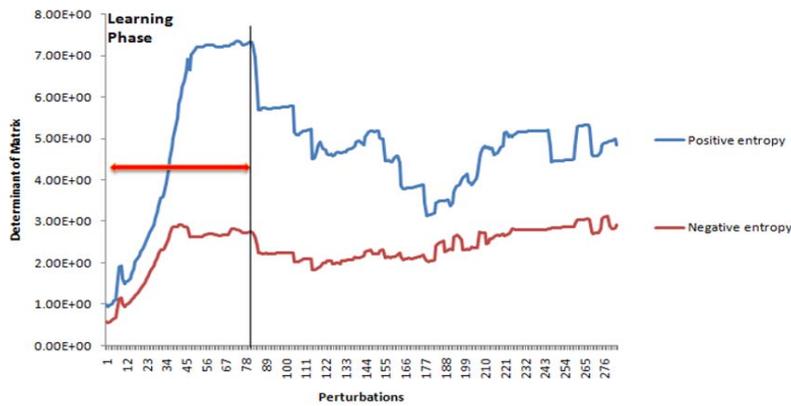

**Fig-2A legends:** In the mixed strategy game, **A** and **B** continuously update their strategy, learning from the opponent. Depth of NE minima (local and global attractors) in TCA cycle phase-space ensures that NE profile always lies low, even under dynamic PE perturbations. First 76 iterations describe learning phase (nascent adaptation); after that **A** and **B** play with a common strategy.

**Fig-2B:**
Title: Description and measurement of biological robustness

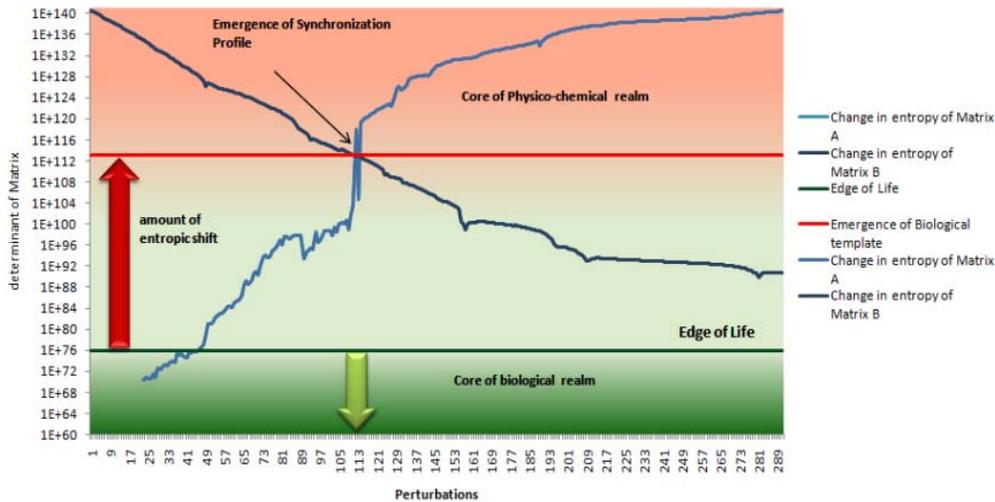

**Fig-2B legends**: 'Core of biological realm' describes the biologically functional domain of TCA cycle; it is characterized by order order transitions, presence of global attractors, SP and NE. 'Core of physico-chemical realm' describes the randomly fluctuating states of physico-chemical environment; characterized by order disorder transitions. Under dynamically varying PE perturbations, TCA cycle leaves its functionally active zone and enters its Edge Of Life(EOL), a zone in phase-space where order order transitions are not consistent. Perturbed further, system attempts to regain SP by clinging to local attractors (that describes the sudden dip in its profile), before dying (viz. entering physico-chemical realm, marked by order disorder transitions). Biological robustness is proportional to 'Amount of entropic shift'; it can be measured in unambiguous numeric terms at a systemic level.

Figure 3:

Title : COPASI and deterministic differential equation based verification

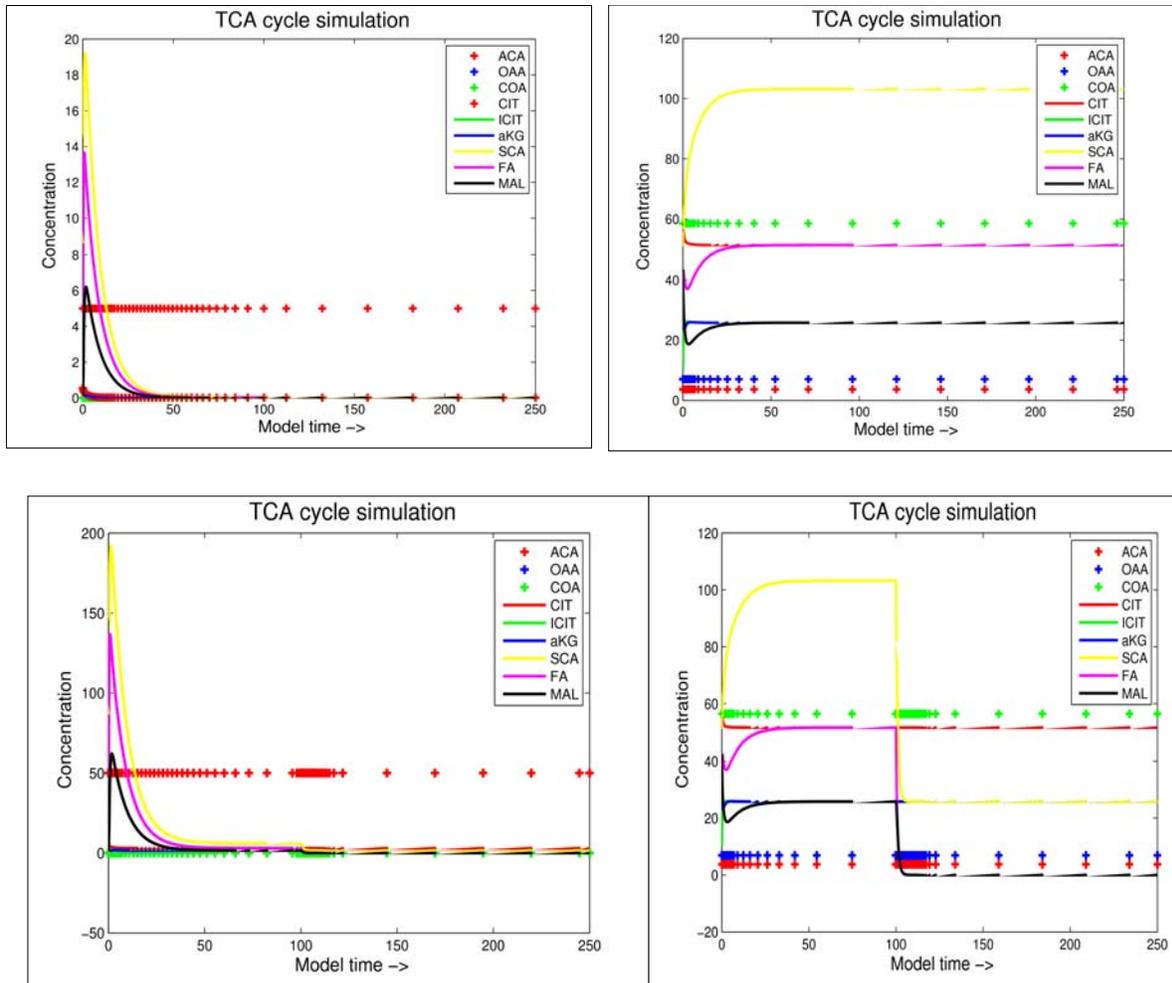

**Figure 4 Legends: Fig-3.A** shows the simulation of TCA cycle when the metabolite concentrations are at biologically functional state('Core of biological realm' in **Fig-2B**). **Fig-3.B** describes these metabolites in 'core of physico-chemical realm' (**Fig-2B**). **Fig-3.A** and **Fig-3.B** findings confirm fully the findings from **Fig-1.A and Fig-1.B**, respectively. **Fig-3.C** and **3.D** shows the simulation of TCA cycle for Biological state and Physico-chemical state respectively. This shows the loss of SP globally in the case of 3.C and loss of SP locally in the case of 3.D. (details in Supplementary material)